\documentclass[aps,12pt,final,notitlepage,oneside,onecolumn,nobibnotes,nofootinbib,superscriptaddress,noshowpacs,centertags]{revtex4}

\usepackage{amsmath,amssymb,epsfig,placeins,subfigure,wrapfig,indentfirst,makeidx}
\usepackage[T1, T2A]{fontenc}
\usepackage[cp1251]{inputenc}
\usepackage[russian,english]{babel}
\usepackage{amsfonts,amssymb}
\usepackage{mathrsfs}
\usepackage{dsfont}
\usepackage[bookmarks=true,pdftex,unicode,plainpages=false]{hyperref}

\input maik2.rty

\begin{document}

\title{The spectrum of the relativistic radiation of electric charges and dipoles in their free falling into a black hole}

\author{A.A. Shatskiy\footnote{shatskiy@asc.rssi.ru}}
\affiliation{The Astro Space Center, Lebedev Physical Institute of
RAS, 84/32 Profsoyuznaya st., Moscow, 117997, Russia}

\author{I.D. Novikov}
\affiliation{The Astro Space Center, Lebedev Physical Institute of
RAS, 84/32 Profsoyuznaya st., Moscow, 117997, Russia}
\affiliation{The Nielse Bohr International Academy, The Nielse
Bohr Institute, Blegdamsvej 17, DK-2100 Copenhagen, Denmark}

\author{L.N. Lipatova}
\affiliation{The Astro Space Center, Lebedev Physical Institute of
RAS, 84/32 Profsoyuznaya st., Moscow, 117997, Russia}

\date{\today}

\begin{abstract}
The free fall of electric charges and dipoles, radial and
freely falling into the Schwarzschild black hole event horizon, was
considered. Inverse effect of electromagnetic fields on the black
hole is neglected. Dipole was considered as a point particle, so
the deformation associated with exposure by tidal forces are
neglected. According to the theorem, "the lack of hair"\, of
black holes, multipole magnetic fields must be fully emitted by
multipole fall into a black hole. The spectrum of electromagnetic
radiation power for these multipoles (monopole and dipole) was
found. Differences were found in the spectra for different
orientations of the falling dipole. A general method has been developed to find
radiated electromagnetic multipole fields for the free falling multipoles
into a black hole (including higher order multipoles
--- quadrupoles, etc.). The electromagnetic spectrum can be
compared with observational data from stellar mass and
smaller black holes.

\end{abstract}

\maketitle

\section{Introduction}
\label{s1}

It is known \cite {Thorn1998, Frolov1998},
black hole has no "Hair", so all multipole moments of the
electromagnetic fields disappear as the system of charges close to
the horizon of the Schwarzschild black hole. Field problem for
static point charge has been solved by Linet  \cite {Linet1976}.
In this problem it has been shown, in particular, that the field
of a point charge close to the field of a charged black hole with
the same charge, when the charge become close to the horizon. So all electric and magnetic multipoles
moments should be radiated, when the charge (or a
system of charges or currents) approach to the horizon of the Shwartzshild
black hole.

Loss in the energy is defined essentially by bremsstrahlung, when monopole's (unit charge) motion is accelerated. This is dipole radiation, since his power is inversely proportional to the $c^3$, and radiated field components are inversely proportional to the $c^2$.

The existence of dipole radiation is unobvious
in the case when the massive dipole fall into black hole, because both charges of dipole are moving and accelerated in the same
direction, and the signs of the charges are opposite. However,
increasing curvature of space leads to the radiation of the dipole
type (see section  \ref{s5}).

Complete analogy can be made with quadrupole radiation of gravitational waves in the orbital motion of the two masses (whose power is inversely proportional to ${c^5}$) in quadrupole approximation. In this case the signs of charges and masses are same, and velocities and accelerations are oppositely directed.

\section{
The law of motion of free falling particle} \label{s2}

Let's consider the radial free falling charge or electric dipole with mass
$ m $ on the Schwarzschild black hole horizon. Schwarzschild metric has the form
\footnote{Here the system of units is selected, in which $ {c = 1}
$ and $ {G = 1} $ --- the speed of light and the gravitational
constant.}:
\begin{equation}
ds^2 = \left(1-\frac{r_g}{r}\right) dt^2 -
\left(1-\frac{r_g}{r}\right)^{-1} dr^2 - r^2\, (d\theta^2 +
\sin^2\theta\, d\varphi^2). \label{1-1}\end{equation} Here $ {r_g =
2M}$ is the radius of the horizon of the Shwartzshild black hole.
As you know, rapidly moving charge radiates. There is no radiation in the local
neighborhood of the charge in their own co-moving frame \footnote{Co-moving freely falling frame of reference is
not a rigid system, so there is always a relative acceleration between its different points, which can be neglected only in the
locally small neighborhood of the charge (in Einstein's falling elevator). There is no radiation in this free falling "elevator"\ from free falling charge (in the approximation of small
deformations of the reference frame in the "elevator"\, during his
fall.)}.

The acceleration of a massive particle \footnote {Mass of
the falling particle assumed to be negligibly small compared with
the mass of the black hole.} is defined by well-known expressions in a rigid system of reference at rest with respect to the black hole (see\cite{Landau1988}, \S 87):
\begin {equation}\frac {du ^ i}
{ds} = - \Gamma ^ i_ {kj} u ^ k u ^ j.\label {1-2} \end {equation}
Here $ u ^ i $ is component of $ {\mbox {4-velocity}} $ and
$ {\Gamma ^ i_ {kj}} $ are Christoffel symbols. We have
for a Schwarzschild
black hole (see \cite {Landau1988}, \S 100, 102) \footnote {Here and dot means differentiation with respect to
time$ t $.}:
\begin{eqnarray}
\Gamma^r_{tt}(r)=\frac{r_g}{2r^2}(1-r_g/r)\, , \quad
\Gamma^r_{rr}(r)=-\frac{r_g}{2r^2(1-r_g/r)}\, ,  \quad {\dot r}_e
=-\left(1-\frac{r_g}{r_e}\right)\sqrt{\frac{r_g/r_e - r_g/r_0}{1-
r_g/r_0}} , \,\,
\label{1-4} \\
v^2(r_e)\equiv ({\dot r}_e)^2\frac{|g_{rr}|}{g_{tt}}=\frac{r_g/r_e
- r_g/r_0}{1- r_g/r_0}\, ,\quad
u^t(r_e)=\frac{\sqrt{1-r_g/r_0}}{1-r_g/r_e}, \quad
u^r(r_e)=-\sqrt{r_g/r_e-r_g/r_0}\, . \,\, \label{1-5}
\end{eqnarray}
Here $ {r_e (t)} $ is the radius of the particle, ${r_0}$ is the
radius at which the falling of the particle has been started and ${v (t)}$ is
a three-dimensional velocity of the particle in the Schwarzschild
coordinates.

The law of motion $ {t (r_e)} $ of the test particle is noticed in many papers (\cite {Frolov1998}, \S 2.4), which is free falling in the Schwarzschild field by radial trajectory from an infinite radius:

\begin{eqnarray}
\frac{t(r_e)}{r_g} = const -
\frac{2}{3}\left(\frac{r_e}{r_g}\right)^{3/2}  -
2\sqrt{\frac{r_e}{r_g}} +
\ln\left[\frac{\sqrt{r_e/r_g}+1}{\sqrt{r_e/r_g}- 1}\right].
\label{2-1}
\end{eqnarray}

We need a similar law for a fall from a finite radius $
r_0 $.

The original formula is \footnote {The similar formula is in the work \cite
{Frolov1998}, but it is
for dependence of falling particle's own time $\tau$ on radius.}:

\begin{eqnarray}
t(r_e) =  \sqrt{r_e(r_0-r_e)(r_0/r_g-1)} +
\frac{(r_0+2r_g)\sqrt{r_0/r_g-1}}{2}\cdot \arccos\left(
\frac{2r_e}{r_0} -
1\right)  + \nonumber \\
+ r_g\cdot\ln \left[ \frac{2\sqrt{r_e(r_0-r_e)(r_0/r_g-1)} +r_0
+r_e(r_0/r_g- 2)}{r_0(r_e/r_g-1)} \right]. \label{2-2}
\end{eqnarray}
So the initial conditions correspond to $ {t {(r_e = r_0)} = 0}
$.

Let's imagine that there are clocks showing Schwarzschild time $ t $ along the entire trajectory of the falling particle. Time ${\Delta t +\Delta t_1}$ passed as long as the
light signal from the clock (which the particle flies by) reaches the observer. The
time interval ${\Delta t}$ is the distance between the particle
(when the particle flies past the clock) and its initial location (at the
radius $ r_0 $). The time interval ${\Delta t_1 = const }$ corresponds to
the distance between the initial radius $ r_0 $ and a distant
observer.

According to  \cite {Landau1988}, \S 101 for period $ {\Delta t}
$ we have
\begin{eqnarray}
\Delta t = r_0 - r_e + r_g\ln \left[\frac{r_0-r_g}{r_e-r_g}\right].
\label{2-3}
\end{eqnarray}
It is assumed here and below, that the distant observer is almost on
the same line, as the particle. Dependence on a small angle
$ {\theta << \pi} $ can be negleted (in linear approximation) between radius vectors of the observer and of the particle in the formula (\ref {2-3}).

Measurement of fall velocity of the particle by a distant observer must be described by time ${T(r_e) \equiv t + \Delta t +
\Delta t_1}$:
\begin{eqnarray}
T(r_e) = \sqrt{r_e(r_0-r_e)(r_0/r_g-1)} +
\frac{(r_0+2r_g)\sqrt{r_0/r_g-1}}{2}\cdot \arccos\left(
\frac{2r_e}{r_0} -
1\right)  + \nonumber \\
+ r_g\cdot\ln \left[ \frac{2\sqrt{r_e(r_0-r_e)(r_0/r_g-1)} +r_0
+r_e(r_0/r_g- 2)}{r_0(r_e/r_g-1)} \right] + r_0 - r_e + r_g\ln
\left[\frac{r_0-r_g}{r_e-r_g}\right] + const.\quad\label{2-4}
\end{eqnarray}
Falling particle reach the horison in infinite time due to the logarithms in (\ref {2-2}), (\ref {2-3}) and (\ref{2-4}).

Let's introduce required notations:

\begin{eqnarray}
a\equiv\sqrt{1-r_g/r_e} ,\quad b\equiv\sqrt{r_g/r_e - r_g/r_0}
,\quad c_1\equiv\sqrt{1-r_g/r_0}\, . \label{abc-1} \end{eqnarray}
We have according to expressions (\ref {2-2}) and (\ref {2-3}):
\begin{eqnarray}
\frac{dT}{dr_e} = \frac{1}{\dot r_e} - \frac{1}{1-r_g/r_e} = -
\frac{b + c_1}{a^2 b},
\label{2-5} \\
\beta (r_e)\equiv\frac{dr_e}{dT} = - \frac{a^2 b}{b + c_1}.
\label{2-6}
\end{eqnarray}

It is important to note, that $ {\beta = 0} $ for $ {r_e = 0} $ and
$ {r_e = r_g} $ (as well as for the quantity $ {\dot r_e} $).

Physical meaning of the $ \beta $ is observable (by a
distant observer) velocity of the falling particle.

\section{Spectral density of radiation
} \label{s4}

We use different method, than in the work  \cite {Zerilli1970, Davis1971, Teukolsky1972, Yakovlev1975, Dymnikova1977},
to calculate the spectral density of the electromagnetic radiation.

The spectral density of the electromagnetic radiation from the
radial current (which flows along $ {\theta_e = 0} $) is given
by (\cite{Landau1988}, \S 66):
\begin{eqnarray}
d{\cal E}_{\rm w} = - \frac{{\bf\hat E_{\rm w}^\theta} {\bf\hat
H_{\rm w}^\varphi}}{4\pi^2}\cdot r^2 \cdot do \cdot d{\rm w}  \,
,\quad do\equiv 2\pi\sin\theta\, d\theta \, . \label{6-1}
\end{eqnarray}
Here $ {{\bf \hat E_ {\rm w} ^ \theta} ({\rm w})} $ and $ {{\bf
\hat H_ {\rm w} ^ \varphi} ({\rm w })} $ are spectral densities
of the physical components of electric and magnetic fields, which
are orthogonal to the direction of wave propagation (at
infinity); $ {{\rm w} \equiv k_t} $ --- the time component of the 4D
null vector of photon $ k_j $, and the components $ k_t $ and $ k_
\theta $ are integrals of motion for each of the emitted photon.

General relativistic generalization of this formula is:

\begin{eqnarray}
d{\cal E}_{\rm w} = - \frac{({\bf{F}_{\rm w}^{\theta
t}}\sqrt{|g_{\theta\theta} g_{tt}|}) ({\bf{F}_{\rm w}^{\theta
r}}\sqrt{|g_{\theta\theta} g_{rr}|})}{4\pi^2}\cdot r^2\cdot do
\cdot d{\rm w}. \label{6-1-2}
\end{eqnarray}
The expression (\ref {6-1-2}) at infinity (far from the black hole)
can be rewritten using that components of
the electric and magnetic fields are equal by magnitude in the electromagnetic wave:

\begin{eqnarray}
d{\cal E}_{\rm w} = - \frac{|{\bf { F}_{\rm w}^{\theta
t}}|^2}{4\pi^2}\cdot r^4\cdot do \cdot d{\rm w} = - \frac{|{\bf {
F}_{\rm w}^{\theta r}}|^2}{4\pi^2}\cdot r^4\cdot do \cdot d{\rm w}.
\label{6-1-4}
\end{eqnarray}
Thus radiated Fourier components of the field $ {{\bf {F} _ {\rm
w} ^ {\theta t}}} $ and $ {{\bf {F} _ {\rm w} ^ {\theta r}} } $ at
infinity must be inversely proportional to the square of the
distance in order to satisfy the condition of energy conservation
of electromagnetic waves.

The total energy radiated to infinity can be obtained either by
integrating over frequency of expression (\ref {6-1-4}), or by
integration of the Poynting vector over time.

Spectral density tensor components $ {{\bf {F} _ {\rm w} ^ {mn}}} $ of the electromagnetic field and the contravariant
components of the electromagnetic field $ {F ^ {mn}} $ are associated
with each other:

\begin{eqnarray}
{\bf { F}_{\rm w}^{mn}} = \int\limits_{-\infty}^{+\infty} F^{mn}
\, \exp{\left[+i {\rm w} t \right]}\, dt,
\label{F1}\\
F^{mn} = \frac{1}{2\pi}\int\limits_{-\infty}^{+\infty} {\bf {
F}_{\rm \Omega}^{mn}}\, \exp{\left[-i \Omega t \right]}\, d\Omega.
\label{F2}
\end{eqnarray}
Thus the spectral densities are obtained from the field components
by the usual Fourier transform.

Because we assume, that beginning of the fall of the particle is the
time $ {t = 0} $ for a distant observer, in this moment the
radiation field must be absent ($ {F ^ {mn} _ {\makebox
{\scriptsize rad.}} = 0 } $ for $ {t <0} $). This corresponds to
half of the sum of the even and odd components of the radiation
field, provided that the components we believe matching in
absolute value.

(${|F^{mn}_{\makebox{\scriptsize
rad.even.}}|=|F^{mn}_{\makebox{\scriptsize rad.odd.}}|=
|F^{mn}_{\makebox{\scriptsize rad.}}|}$).

Then the Fourier transformation of (\ref {F1})-(\ref {F2}) for the
radiation fields can be rewritten as:

\begin{eqnarray}
{\bf { F}_{\rm w}^{mn}} = \int\limits_{0}^{\infty} F^{mn} \,
\frac{1}{2}\left(\sin\left[{\rm w} t \right] + \cos\left[{\rm w} t
\right]\right)\, dt = \int\limits_{0}^{\infty} F^{mn} \,
\frac{1}{\sqrt{2}}\cos\left[{\rm w} t -\frac{\pi}{4}\right]\, dt,
\label{6-1-3-0}\\
F^{mn} = \frac{1}{4\pi}\int\limits_{0}^\infty {\bf { F}_{\rm
\Omega}^{mn}}\, \left(\sin\left[\Omega t \right] +
\cos\left[\Omega t \right]\right)\, d\Omega =
\frac{\sqrt{2}}{4\pi}\int\limits_{0}^\infty {\bf { F}_{\rm
\Omega}^{mn}}\, \cos\left[\Omega t -\frac{\pi}{4}\right ]\,
d\Omega. \label{6-1-3}
\end{eqnarray}
In these formulas the sine corresponds to the odd component and
the cosine corresponds to the even component of the radiation.

\section{Radiation of a charge}
\label{s5}

We use the second pair of
general relativistic Maxwell equations to find components of the spectral density $ {{\bf {F} _ {\rm
w} ^ {mn}}} $ of the electromagnetic field \footnote {The sum over $ e
$ is over all charges (for a dipole --- over the first and the
second charges).}  (see  \cite {Landau1988}, \S 90):

\begin{eqnarray}
\partial_i \left(\sqrt{-g} F^{in}\right) = 4\pi\sqrt{-g} j^n \, ,\quad
j^n (r, t, r_{0}) = \sum\limits_e \frac{q_e\, u_e^n}{\sqrt{-g}\,
u_e^t}\, \delta \left[r - r_e(t)\right]\cdot \delta (\theta) \cdot
\delta (\varphi) \, . \label{6-3}
\end{eqnarray}
Here $ {\sqrt {-g} = r ^ 2 \sin \theta} $ and $ {\delta \left [r
- r_e (t) \right]} $ is delta function with a singularity at the
point, where it is the charge at time $ t $.

If $ {r> r_0} $, the right side of equation (\ref {6-3}) is always
zero (4-vector current $ {j ^ n = 0} $), so we obtain following result by integrating of component
$ j ^ t $ of equation (\ref {6-3}) over $\theta$  in $ {[0, \, \theta]} $:

\begin{eqnarray}
\int\limits_0^\theta \partial_r  \left(\sqrt{-g} F^{rt}\right)\,
d\theta + \sqrt{-g} F^{\theta t} = 0 \label{6-4}
\end{eqnarray}
We rewrite (\ref {6-4}) in the linear approximation in the small angle $ {\theta << \pi} $ as \footnote {Dependence of the field on
the angle $ \theta $ corresponds to the known relativistic
expression of the work  \cite {Ross1971} or \cite {Landau1988},
\S 67.}:

\begin{eqnarray}
r^2 F^{\theta t} = -\frac{\theta}{2} \partial_r \left[ r^2 F^{rt}
\right] = \frac{\theta}{2} \partial_r \left[ r^2 F_{rt} \right] =
\frac{\theta}{2} \partial_r \left[ r^2 (\partial_r A_t -
\partial_t A_r) \right] \label{6-5-0}
\end{eqnarray}
Here $ A_i $ is a 4-vector of the electromagnetic field of a
falling charge in the stationary reference frame of a distant
observer. Falling particle moves with a 4-speed $ U ^ i $
relatively distant observer. Therefore expression (\ref {2-6}) helps us to find the form of $U^i$ \footnote {We neglect the
curvature of space-time at the point of a distant observer in the expression (\ref {6-6-0}).}:
\begin{eqnarray}
U^i \to \left\{\frac{1}{\sqrt{1-\beta^2}},\,
\frac{\beta}{\sqrt{1-\beta^2}},\, 0,\, 0 \right\} \label{6-6-0}
\end{eqnarray}
Here and below the arrow indicates the limit as $ {r / r_g \to
\infty} $.

We use two invariants to find the components $ A_t $ and $ A_r $
in the frame of distant observer : $ {inv_1 = (A_i U ^ i) = \tilde
A_t} $ and $ {inv_2 = (A_i A_j g ^ {ij}) = (\tilde A_t) ^ 2} $.
Here $ {\tilde A_t} $ is the only non-zero component of
4-vector $ {\tilde A_i} $ of the electromagnetic field in the
free falling frame reference associated with the particle. The
expression for $ {\tilde A_t} $ is known from work \cite
{Linet1976}:

\begin{eqnarray}
\tilde A_t = \frac{q_e}{r r_e} \left[\frac{(r-M)(r_e-M)-M^2
\cos\theta}{\sqrt{(r-M)^2 + (r_e-M)^2 - M^2
-2(r-M)(r_e-M)\cos\theta +M^2\cos^2\theta}} + M \right]
\label{6-Linet}
\end{eqnarray}
Here $ {M = r_g / 2} $ is the mass of the Schwarzschild black
hole. In the linear approximation respect to the angle $ \theta $
we obtain \footnote {Expression (\ref {6-Linet}) depends on the
small angle $ \theta $ only in the quadratic approximation.}:

\begin{eqnarray}
\tilde A_t{}_{(\theta =0)} = \frac{q_e(r-r_g)}{r(r-r_e)} \to
\frac{q_e}{r} \label{6-5-2}.
\end{eqnarray}
This shows that the required asymptotic of capacity has form of the Coulomb field in the free falling co-moving frame of reference. This essential and logical result is a consequence of the local inertial frame of reference in the falling "Einstein's elevator."

Then we obtain the required components with omitting all intermediate calculations:
\begin{eqnarray}
A_t = U_t \tilde A_t \to \frac{q_e}{r}\cdot
\frac{1}{\sqrt{1-\beta^2}},
\label{6-6-1}  \\
A_r = U_r \tilde A_t \to -\frac{q_e}{r}\cdot
\frac{\beta}{\sqrt{1-\beta^2}} \label{6-6-2}.
\end{eqnarray}
Expression (\ref {6-5-0})
can be written in the limit $ {r / r_g \to \infty} $ as \footnote {A member $ {\partial_r A_t} $ has an
asymptotic $ {q_e / r ^ 2} $, so it decreases faster at infinity
than a member $ {\partial_t A_r} $ (which has asymptotics $ {q_e / r}
$).}:

\begin{eqnarray}
F^{\theta t} \to -\frac{\theta}{2r^2}\cdot \partial_r \left[ r^2
\partial_t A_r \right] \to -\frac{\theta q_e}{2r^2}\cdot Z \,
,\quad Z\equiv\frac{d}{dt}\left[ \frac{\beta}{\sqrt{1-{\beta}^2}}
\right] \, . \label{6-6-3}
\end{eqnarray}
Here $ {F ^ {\theta t}} $ is the tangential electric
field of the radiation of a point charge when it is in the distance $ {r_e} $ from the black hole. It should be noted that this
field is dipolar, since it is proportional to $ r_g $ and inversely proportional to the square of the speed of light.

We will get from (\ref {6-6-3}) using the derivative with respect to time:

\begin{eqnarray}
Z = \frac{\dot r_e}{\left(1-{\beta}^2\right)^{3/2}} \cdot
\frac{\partial {{\beta}}}{\partial r_e} = -\frac{r_g a^2 (b+c_1)
[a^2 c_1 - 2b^2 (b+c_1)]}{2c_1 r_e^2 [(b+c_1)^2 - a^4 b^2]^{3/2}}
\label{6-6-4}
\end{eqnarray}

\begin{figure*}
\includegraphics[width=0.95\textwidth, height=0.3\textheight]{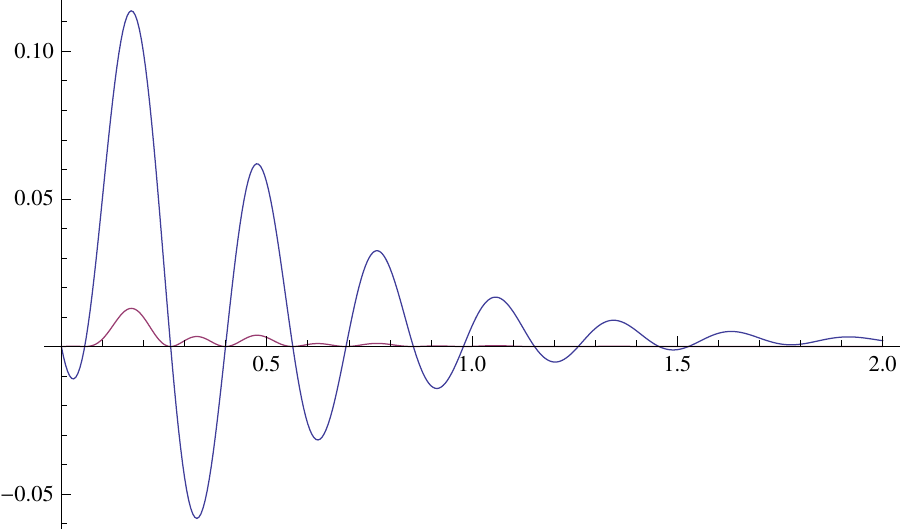}
\caption{{
{\bf Radial fall of a point charge $ q_e $ from radius $ {r_0 = 5r_g} $ into the black hole.} \\
Curve with variable sign and greater amplitude is the spectral density
(Fourier transform) of the radiation field
${{\bf F_{\rm w}^{\theta t}}}$ of the argument ${({\rm w} r_g)}$ in units ${\left[q_e\theta /r^2 \right]}$.\\
Curve with fixed sign and lower amplitude is the square of  ${{\bf
F_{\rm w}^{\theta t}}}$, which is the power spectrum of
electromagnetic radiation (\ref{6-1-4}). }} \label{R1}
\end{figure*}

We obtain by substituting (\ref {6-6-3}) and (\ref {6-6-4}) in the Fourier
transform (\ref {6-1-3-0}):

\begin{eqnarray}
{\bf {F}_{\rm w}^{\theta t}} = \int\limits_{0}^{\infty}
\frac{\cos\left[{\rm w} t -\frac{\pi}{4}\right]}{\sqrt{2}} \,
F^{\theta t}\, dt = -\frac{\theta q_e}{r^2}
\int\limits_{0}^{\infty} \frac{\cos\left[{\rm w} t
-\frac{\pi}{4}\right]}{2\sqrt{2}}\,
\partial_t \left[ \frac{\beta}{\sqrt{1-{\beta}^2}} \right] \, dt
\label{6-7}
\end{eqnarray}
We obtain by replacing in (\ref {6-7}) integration variable: $ {dt = dr_e /
{\dot r_e}} $, according to (\ref {abc-1}) for the spectral density
of the radiation field:

\begin{eqnarray}
{\bf {F}_{\rm w}^{\theta t}} = -\frac{\theta q_e}{r^2}
\int\limits_{0}^{\infty} \frac{{\rm w}\, \sin [{\rm w} t
-\frac{\pi}{4}]\, \beta}{2\sqrt{2}\sqrt{1-{\beta}^2}} \, dt =
\frac{\theta q_e}{r^2} \int\limits_{r_g}^{r_0} \frac{{\rm w}\,
c_1\, \sin \left[{\rm w} t(r_e)
-\frac{\pi}{4}\right]}{2\sqrt{2}\sqrt{(b+c_1)^2 - a^4 b^2}} \,
dr_e \label{6-7-1}
\end{eqnarray}
We used here the integration by parts and we noted that
$ {\beta = 0} $ in points $ {r_e = 0} $ and $ {r_e = r_g} $. The
function $ {t (r_e)} $ is given by expression (\ref {2-2}).

Results obtained using (\ref {6-7-1}) are shown in Figure  \ref
{R1}. The spectral density of the radiation field of a
charge is shown below. Here we use posibility of integration (\ref
{6-7}) in parts, according to the expression (\ref {6-6-2}) and
(\ref {6-7-1}).
\begin{eqnarray}
{\bf {F}_{\rm w}^{\theta t}} = \frac{\theta {\rm w}}{r}\cdot{\bf
A_{r{\rm w}}} \label{6-7-2}
\end{eqnarray}
Here $ {\bf A_ {i {\rm w}}} $ is the Fourier transform of the
retarded vector potential for the field of the charge. It
corresponds to the usual definition of the Fourier components of
the radiation field ($ {{\bf | \hat E_ {\rm w} ^ \theta |} = {\rm
w} \sin \theta | \vec {\bf A} _ {\rm w} |} $ --- see  \cite
{Landau1988}, \S 66), and it is possible by
ability of integration by parts in (\ref {6-7}). But dipole field representation in the form (\ref
{6-7-2}) is not possible, as it will be shown in the next
section.

\section{Dipole radiation}
\label{s6}

The value of the dipole is defined as: $ {d_ {0} \equiv q_e l_0 =
const} $. We assume the dipole as point particle ($ {l_0 <<< r_g} $) and neglect the
deformation related to effects of tidal forces.
We denote the radius of the dipole location: $ {r_d \equiv (r_1 +
r_2) / 2} $, then respectively replace $ {r_e \to r_d} $ in all expressions for dipoles.

\subsection{
Transverse orientation of the dipole} \label{s6-2}

First we choose the orientation of the dipole across the radius
(the simplest case). Then $ {l_0 = r_d \Delta \theta_d} $, so the
expression (\ref {6-6-3}) for the tangential electric field of the
dipole radiation in this case will be rewritten as \footnote {Formula
(\ref {7-6}) differs from the corresponding formula (\ref {6-6-3})
only by coefficient and factor $ {r_d} $ in the denominator of the
integrand.}:

\begin{eqnarray}
F^{\theta t}_{d_\perp} = \Delta\theta_d \cdot \frac{\partial
F^{\theta t}}{\partial \theta} = \frac{- d_0 Z}{2 r_d r^2} =
\frac{d_0}{r^2}\cdot \frac{r_g a^2 (b+c_1) [a^2 c_1 - 2b^2
(b+c_1)]}{4 c_1 r_d^3 [(b+c_1)^2 - a^4 b^2]^{3/2}} \label{7-6}
\end{eqnarray}
The function $ {F ^ {\theta t} _ {d_ \perp} {} _ {(t, r)}} $ is
the tangential electric field of the dipole radiation at the time
of its location at a distance $ {r_d} $ from the black hole (for
transverse orientation).

\begin{figure*}
\caption{{\bf Radial fall of a dipole from radius $ {r_0 = 5r_g} $
on the black hole.} \\ For transversely oriented dipole
(top): variable sign curve with greater amplitude is the spectral
density (Fourier transform) of the radiation field ${{\bf {F}_{\rm
w{d_\perp}}^{\theta t}}}$ of the argument ${({\rm w} r_g)}$, in
units of ${\left[d_0/(r_g r^2)\right]}$. Fixed sign curve with
lower amplitude is the square of ${{\bf {F}_{\rm
w{d_\perp}}^{\theta t}}}$, which is proportional to the power
spectrum (\ref {6-1-4}) of electromagnetic radiation. \label{R2}}
\subfigure{\includegraphics[width=0.95\textwidth,
height=0.28\textheight]{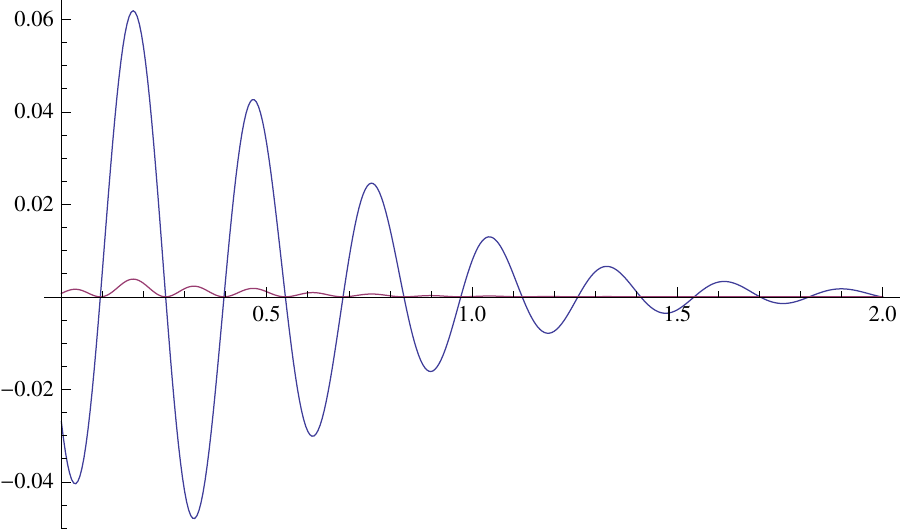}}
\subfigure{\includegraphics[width=0.95\textwidth,
height=0.28\textheight]{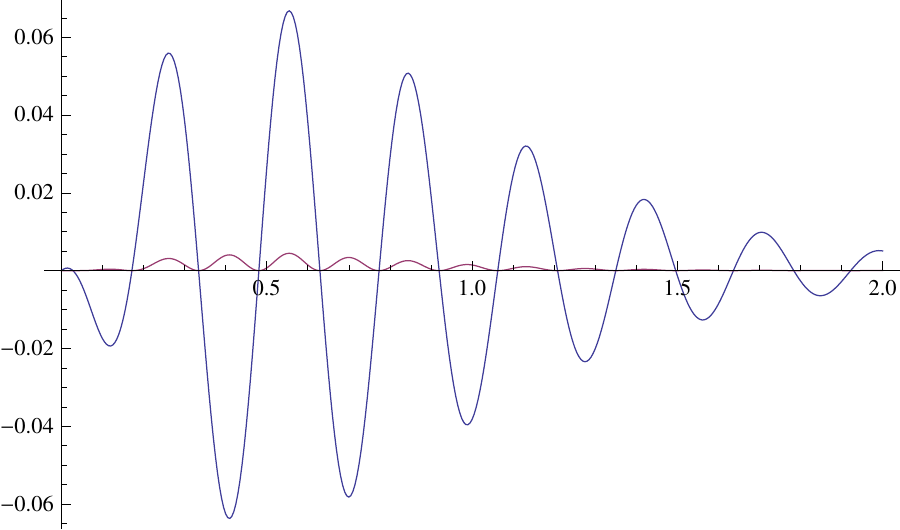}} \caption{{ \textbf{The same thing for
longitudinally oriented dipole is ${{\bf {F}_{\rm
w{d_\parallel}}^{\theta t}}}$, in units of ${\bf\left[d_0\theta /(r_g
r^2)\right]}$.} \label{R3} }}
\end{figure*}

$\quad$ \newpage

So the spectral density of the radiation field of a
dipole takes the form in the transverse orientation:

\begin{eqnarray}
{\bf F_{{\rm w}d_\perp}^{\theta t}} = \int\limits_{0}^{\infty}
\frac{F^{\theta t}_{d_\perp}\, \cos\left[{\rm w} t
-\frac{\pi}{4}\right]}{\sqrt{2}}\, dt = \frac{d_0}{r^2}
\int\limits_{r_g}^{r_0} \frac{r_g\, \cos [{\rm w}t(r_d)
-\frac{\pi}{4}] \, (b+c_1) [a^2 c_1 - 2b^2 (b+c_1)]}{4\sqrt{2} b
r_d^3 [(b+c_1)^2 - a^4 b^2]^{3/2}} \, dr_d \quad \label{7-7}
\end{eqnarray}
Here the variable of
integration was replaced with ${dt = dr_d/{\dot r_d}}$, as in the expression (\ref {6-7-1}).

It is important to note that
the dipole radiation is independent of $ \theta $ in the linear approximation by
the angle $ {\theta << \pi} $
in the case of the transverse orientation
(since
the radiation pattern of the dipole is located just in the
transverse direction).

\subsection{
Longitudinal orientation of of the dipole} \label{s6-1}

Now let's choose the orientation of the dipole along the radius. Length of
the dipole can be calculated using following formula taking into account the curvature of space:

\begin{eqnarray}
l_0\approx [r_2(t)-r_1(t)]/\sqrt{1-r_g/r_d} \label{7-1}
\end{eqnarray}
Hence it follows
\begin{eqnarray}
\Delta r_d\equiv r_2(t)-r_1(t) = a\cdot l_0 \, , \quad \Delta r_0
= c_1\cdot l_0 \label{7-2}.
\end{eqnarray}
Here $\Delta$ denotes the change at constant time $t$, so differential must be zero:

\begin{eqnarray}
dt = \frac{\partial t}{\partial r_0}\cdot \Delta r_0 +
\frac{\partial t}{\partial r_d}\cdot \Delta r_d = 0 \label{7-3-1}.
\end{eqnarray}
Expression (\ref {6-6-3}) for a dipole can be written as:

\begin{eqnarray}
F^{\theta t}_{d_\parallel} = \Delta F^{\theta t} \equiv
\frac{\partial F^{\theta t}}{\partial r_0}\cdot \Delta r_0 +
 \frac{\partial F^{\theta t}}{\partial r_d}\cdot \Delta r_d = \left[
\frac{\partial F^{\theta t}}{\partial r_0} - \frac{\partial
F^{\theta t}}{\partial r_d}\cdot \frac{\partial t}{\partial r_0}
\cdot {\dot r_d} \right]\cdot c_1 \cdot l_0 \label{7-3}.
\end{eqnarray}

Similarly to the case of the transverse orientation of the dipole
the spectral density of the radiation field for the dipole in the
longitudinal orientation takes the form:

\begin{eqnarray}
{\bf {F}_{\rm w{d_\parallel}}^{\theta t}} =
\int\limits_{0}^{\infty} \frac{F^{\theta
t}_{d_\parallel}\,\cos\left[{\rm w} t
-\frac{\pi}{4}\right]}{\sqrt{2}}\, dt = \frac{d_0 \theta}{r^2}
\int\limits_{r_g}^{r_0} \frac{c_1\,\cos\left[{\rm w} t(r_d)
-\frac{\pi}{4}\right]}{2\sqrt{2}} \cdot \left[ \frac{\partial
Z}{\partial r_0} - {\dot r_d} \cdot \frac{\partial t}{\partial
r_0} \cdot \frac{\partial Z}{\partial r_d} \right] \,
\frac{dr_d}{\dot r_d} \quad \label{7-4}.
\end{eqnarray}
The derivatives necessary for calculation of the quadrature (with regard to the expression (\ref {6-6-4})) are given in Appendix (see  (\ref
{A-1}), (\ref {A-2}) and (\ref {A-3})).

Plots of the power spectrum for the transverse and longitudinal
dipole orientations are shown in Figure  \ref {R2} and Figure  \ref
{R3}, respectively.

\section{
Discussion and conclusions} \label{s7}

As shown in previous sections, the radiation both falling
monopole and falling dipole ((\ref {7-6}) and (\ref {7-3})) are
dipolar, because radiated field components are proportional to $
r_g $  (\ref {6-6-4}), that is inversely proportional to
$ c ^ 2 $.

The total energy of the radiation received by integration over
time of the Poynting vector is $ {{\cal E} = \frac {1} {4 \pi} \int
\limits (F ^ {\theta t}) ^ 2 r ^ 4 \, do \, dt} $. It is equal $ {{\cal E} ^ e \approx 0.0016 q_e ^ 2/r_g} $
for the monopole and $ {{\cal E} ^ d \approx 0.0007 d_0 ^ 2/r_g ^ 3} $ for
the dipole, taking into
account the angular distribution of power (see  \cite
{Ross1971} or \cite {Landau1988}, \S 67). In
addition, the energy of the
radiation exceeds their rest energy for the falling electrons in the field of black holes
with a mass less than $ 10 ^ {14} {g} $. It speaks about the
infidelity of above calculations for radiation in the classical
theory for primordial black holes (with a mass less than $ {10 ^
{14}} $ g). In these cases it is necessary to use other theory (quantum
gravity).

Found spectrum of dipole radiation depends on the orientation of
the dipole.

Wavelength characteristic $ \lambda_m $ depends on the specific value of $ {r_0} $ and on the orientation of the dipole at the maximum of the
radiation. $ \lambda_m $ is approximately $ {20 r_g} $, when $
{r_0 = 5r_g} $. Therefore it's possible to
observe this radiation near the maximum for a relatively small
black hole (with a mass $ {M <\sim M_ \odot} $). If you'll try to
observe this radiation at shorter wavelengths, spectral power decreases approximately as $ {{\rm w} ^ {-4}} $ in the local maximum .

Since this radiation has a spectrum characteristic, it can be
registered for the rare cases of falling magnetized planets (or
asteroids) on black holes of stellar mass or even in rare cases,
when pulsars fall into the black hole.

\section*{Aknowledgments}
\label{s8}

We are particularly grateful to K.A. Bronnikov for many useful
discussions on the subject and for valuable comments.

This work was supported by RFBR, project codes: 12-02-00276-a,
11-02-00244-a, 11-02-12168-ofi-m-2011, Scientific
School-2915.2012.2 "Formation of large-scale structure of the
Universe and cosmological processes" Programme "Nonstationary
Phenomena in the objects of the universe 2012"\, and the Federal
Target Program "Scientific and pedagogical Staff of Innovative
Russia 2009-2013" \, 16.740.11.0460.

\newpage
\appendix

\subsection{
APPENDIX: necessary for the integration of quadrature (\ref {7-4})
expressions } \label{A1}

We denote $ {x \equiv r_g / r_d} $, $ {y \equiv r_g/r_0} $ and
give the necessary expressions:

\begin{eqnarray}
\frac{\partial Z}{\partial r_0} &=&\frac{(x-1)^2 x^2 y^2}{4 r_g^2
(1-y)^{3/2} \sqrt{x-y}
   \left[1-x^3+x^2 (y+2)-2 x y+2 \sqrt{(1-y)(x-y)}-y\right]^{5/2}}\nonumber\\&\times& \left[
   2x^4 \sqrt{x-y}+x^3 \left(2 \sqrt{1-y}-7\sqrt{x-y}\right)\right.\nonumber\\
&+&\left. 2 \left(x^2-2 x-15\right)
y^2\left(\sqrt{x-y}+\sqrt{1-y}\right)+x^2 \left(9 \sqrt{x-y}-14
   \sqrt{1-y}\right)\right.\nonumber\\
&+&\left. y \left\{ -x^3 \left(\sqrt{x-y}+2\sqrt{1-y}\right)+x^2
\left(2 \sqrt{1-y}-\sqrt{x-y}\right)+x
   \left(29 \sqrt{x-y}+42 \sqrt{1-y}\right)\right.\right.\nonumber\\
&+&\left.\left.37 \sqrt{x-y} + 22\sqrt{1-y}\right\} -x \left(29
\sqrt{x-y}+18 \sqrt{1-y}\right) -
   7\sqrt{x-y}-2 \sqrt{1-y} \right]
\label{A-1}
\end{eqnarray}
\begin{eqnarray}
\frac{dt}{dr_0}=\frac{1-(x+1)y/(2x) +
y^2}{\sqrt{(1-y)(x-y)}}+\frac{\sqrt{1-y}(2y+1)}{2
\sqrt{x-y}}+\frac{3\arccos\left(2y/x-1\right)}{4 \sqrt{y(1-y)}}
\label{A-2}
\end{eqnarray}
\begin{eqnarray}
\frac{\partial Z}{\partial r_d} &=&\frac{x^3 \left[ (2-y) y
(x-y)+2 \sqrt{(1-y)(x-y)} + 1-y \right]^{-1/2} }{4 r_g^2
\sqrt{(1-y)(x-y)}
   \left[(2-x) x (x-y) + 2\sqrt{(1-y)(x-y)} + 1-y \right]^2}\nonumber\\
&\times& \left[2 x^6 \sqrt{x-y}-x^5 \left(17 \sqrt{x-y}+18
   \sqrt{1-y}\right)+x^4 \left(8 \sqrt{x-y}-26 \sqrt{1-y}\right)\right.\nonumber\\
&-&\left. 4 x^3\left(25 \sqrt{x-y}+14 \sqrt{1-y}\right)+16
\left(x^2+4 x-3\right)
   y^3 \left(\sqrt{x-y}+\sqrt{1-y}\right)\right.\nonumber\\
&+&\left.x^2 \left(62 \sqrt{x-y}+78\sqrt{1-y}\right)+ 2 y^2
\left\{ x^4
   \left(\sqrt{x-y}+\sqrt{1-y}\right)-x^3 \left(25 \sqrt{x-y}+29
   \sqrt{1-y}\right)\right.\right.\nonumber\\
&-&\left.\left.x^2 \left(79 \sqrt{x-y}+91 \sqrt{1-y}\right)+x
   \left(33 \sqrt{x-y}+61 \sqrt{1-y}\right)+22 \sqrt{x-y}+10\sqrt{1-y}\right\}\right.\nonumber\\
&+&\left.y \left\{ -x^5 \left(\sqrt{x-y} + 2\sqrt{1-y}\right)+12
x^4 \left(3 \sqrt{x-y}+5
   \sqrt{1-y}\right)\right.\right.\nonumber\\
&+&\left.\left.\left.2 x^3 \left(50 \sqrt{x-y} +71
\sqrt{1-y}\right)
   -4x^2 \left(3 \sqrt{1-y}-26 \sqrt{x-y}\right)-x \left(151
   \sqrt{x-y}+104 \sqrt{1-y}\right)\right.\right.\right. \nonumber\\
&+&\left. \left. 8 \sqrt{x-y}+12 \sqrt{1-y}\right\} +x
   \left(17 \sqrt{x-y}-10 \sqrt{1-y}\right)-4 \sqrt{x-y} \right]
\label{A-3}
\end{eqnarray}
It is interesting that all these expressions are singular at the
point $ {r_d = r_0} $, but their combination in the formula (\ref
{7-4}), is finite.

\end{document}